\documentclass[useAMs,a4paper]{mn2e}
\usepackage{savesym}
\usepackage{graphicx}
\expandafter\let\csname equation*\endcsname\relax
  \expandafter\let\csname endequation*\endcsname\relax 
\usepackage{subfig}
\usepackage{amsmath}
\usepackage{amssymb}
\usepackage{verbatim}
\usepackage[yyyymmdd,hhmmss]{datetime}
\usepackage{array}
\usepackage{times}
\usepackage[total={17.8cm,24.0cm},centering]{geometry} 
\usepackage{color}

\newcommand{\beq}{\begin{equation}}
\newcommand{\eeq}{\end{equation}}

\newcommand\T {{\widetilde {T}}}

\title[TDE discs are larger than they seem]{Tidal disruption event discs are larger than they seem: removing systematic biases in TDE X-ray spectral modelling }
\author [Andrew Mummery]{Andrew Mummery\thanks{E-mail:
andrew.mummery@physics.ox.ac.uk}
\\
Oxford Astrophysics, Denys Wilkinson Building, Keble Road, Oxford, OX1 3RH, United Kingdom}
\begin{document}

\date{}

\pagerange{\pageref{firstpage}--\pageref{lastpage}} \pubyear{2021}

\maketitle

\label{firstpage}

\begin{abstract} 
The physical sizes of TDE accretion discs are regularly inferred, from the modelling of the TDEs X-ray spectrum as a single temperature blackbody, to be smaller than the plausible event horizons of the black holes which they occur around -- a clearly unphysical result. In this Letter we demonstrate that the use of single-temperature blackbody functions results in the systematic underestimation of TDE accretion disc sizes  by as much as an order-of-magnitude. In fact, the radial `size' inferred from fitting a single temperature blackbody to an observed accretion disc X-ray spectrum does not even positively correlate with the physical size of that accretion disc.  We further demonstrate that the disc-observer inclination angle and absorption of X-ray photons may both lead to additional underestimation of the radial sizes of TDE discs, but by smaller factors. To rectify these issues we present a new fitting function which accurately reproduces the size of an accretion disc from its $0.3-10$ keV X-ray spectrum. Unlike traditional approaches, this new fitting function does not assume that the accretion disc has reached a steady state configuration, an assumption which is unlikely to be satisfied by most TDEs. An XSPEC implementation of this new fitting function is available at github.com/andymummeryastro/TDEdiscXraySpectrum.
\end{abstract}

\noindent
%Complied at \today\ \currenttime\ .

\section{introduction} 
In recent years the number of tidal disruption events (TDEs) observed at soft X-ray energies has greatly increased, with a current population of approximately 20 sources. X-ray bright TDEs represent a particularly interesting TDE sub-population, as physical parameters of a TDE system can be inferred from the modelling of their X-ray spectral energy distribution.

A common practice in the TDE literature is to fit the observed X-ray spectrum of a TDE with a single temperature blackbody function (for example, some recent papers: Brown {\it et al}. 2017,  Holoien {\it et al}. 2018, van Velzen {\it et al}. 2019, Wevers {\it et al}. 2019, Stein {\it et al}. 2020, Cannizzaro {\it et al}. 2021,  Hinkle {\it et al}. 2021). Modelling TDE X-ray spectra in this manner allows  two parameters to be inferred: a temperature $T_{\rm bb}$, and `emission radius' $R_{\rm bb}$ (eq. \ref{planck}). The inferred size of the X-ray emitting region is a parameter of interest as, assuming that this emission results from a disc which extends down to the ISCO, it can be used as an estimate for the TDEs central black hole mass, $R_{\rm bb} \simeq R_{\rm ISCO} \propto GM_{\rm BH}/c^2$. The black hole mass at the centre of a TDE is an extremely important physical parameter, as it strongly correlates with the peak luminosity of the disc which forms in the aftermath of a TDE (Mummery \& Balbus 2021a). This luminosity scale in turn controls the observed properties of a particular TDE (Mummery 2021b). Therefore, the radial size of an X-ray bright TDEs disc could, if measured correctly, be used to understand more general properties of the TDE.

A characteristic radial scale of a TDE system is the Schwarzschild radius of its central black hole, $R_S \equiv 2GM_{\rm BH}/c^2$. The mass of a central black hole can be inferred from a number of galactic scaling relationships, and thus these two radii ($R_{\rm bb}$ and $R_S$) can be compared. A surprising -- to say the least -- result of a number of works on X-ray bright TDEs is that the inferred X-ray emission radius $R_{\rm bb}$ is often found to be significantly smaller than the event horizon of the TDEs central black hole (e.g. van Velzen {\it et al}. 2019, Wevers {\it et al}. 2019, Stein {\it et al}. 2020, Cannizzaro {\it et al}. 2021,  Hinkle {\it et al}. 2021). As an explicit example, Wevers {\it et al}. (2019, their Figure 11) find that 13 of 20 X-ray bright TDEs have $R_{\rm bb}/R_S < 1$. This is an unphysical result. 

Thus far { three} possible explanations have been put forward for this `too small emission radius' problem. These are, (i) that TDE discs are highly inclined with respect to the observer (e.g. Stein {\it et al}. 2020),  (ii) that TDE discs are obscured from the observer by other TDE debris (e.g. Wevers {\it et al}. 2019), { or (iii) that disc photons are comptonised before reaching the observer (Saxton {\it et al}. 2021)}. In this work we put forward { an additional}, as of yet unconsidered, explanation for these spurious results: the radius inferred from fitting a single temperature blackbody to an accretion disc spectrum does not correlate with any physical radial scale of the accretion disc. The reason for this is rather straightforward: the Wien tail of a multi-temperature colour-corrected accretion disc spectrum (the part of a TDE spectrum probed by an X-ray telescope) does not have the functional form of a single temperature blackbody. Rather, the correct functional form has multiple correction factors which modify both the amplitude and energy dependence of the observed spectrum (equation \ref{MB}). These modifications result in the systematic underestimation of TDE disc sizes by blackbody modelling. The correction factor from this effect for typical TDE parameters can be as high as an order of magnitude. 

In sections 2 and 3 of this paper we analytically and numerically demonstrate this systematic issue with single temperature blackbody modelling. In section 4 of this paper we examine { two of the} existing explanations for these small emission radii. We demonstrate that highly inclined disc systems can have inferred radii which are a factor $\sim 3$ smaller than those of face-on discs. Finally, we demonstrate that discs with high levels of neutral absorption, as is expected from TDE systems with large initial bolometric luminosities (Metzger \& Stone 2016), also result in TDE radii being underestimated.  The magnitude of this obscuration correction factor depends on the density of absorbing material, but can { also be significant} for the most obscured systems. 
\section{Modelling TDE X-ray spectra}
A common practice in the TDE literature is to fit the observed X-ray spectrum of a TDE with a single temperature blackbody function, which has the following functional form:  
\begin{multline}\label{planck}
F_\nu(R_{\rm bb}, T_{\rm bb}) = \pi \left({R_{\rm bb} \over D}\right)^2 B_\nu(T_{\rm bb}) \\ = 
{2\pi h \nu^3 \over c^2} \left({R_{\rm bb} \over D}\right)^2 {1 \over \exp\left({h\nu / kT_{\rm bb}}\right) - 1} .
\end{multline}
There are three primary simplifications present in this model which result in large deviations from physical reality. Firstly, this model assumes a spherical emission surface, whereas TDE soft X-ray photons will result from an accretion disc. Further, this model assumes that the TDE emission surface evolves with only a single temperature, when in reality the disc temperature will vary with disc radius. Third, this model neglects the effects of so-called disc `colour-correction factors', which model disc {opacity} effects (Done {\it et al}. 2012). Colour-correction factors have large effects on the properties of disc spectra observed in their Wien tail. 

In reality the observed thermal ``soft state'' emission from an accretion disc is given by the following expression 
\beq\label{fullmodel}
F_\nu =  \frac{1}{D^2}\iint_{\cal S} {f_\gamma^3 f_{\rm col}^{-4} B_\nu (\nu/f_\gamma, f_{\rm col} T)} ~\text{d}b_1\text{d} b_2.
\eeq
Here ${\cal S}$ is the surface of the disc, and $b_1$ and $b_2$ are cartesian {image plane} photon impact parameters. The factor $f_\gamma$ is the photons energy-shift factor, defined as the ratio of the observed photon frequency $\nu$ to the emitted photon frequency $\nu_{\rm emit}$, $f_\gamma \equiv \nu/\nu_{\rm emit}$. Here $T$ is the temperature of the disc, a function of both disc radius and time $T(r,t)$. Finally, $f_{\rm col}$ is the `color-correction' factor, which is included to model disc {opacity} effects. This correction factor generally takes a value $f_{\rm col} \sim 2.3$ for typical TDE disc temperatures. For a more detailed derivation and discussion of this expression see Section 2 of Mummery \& Balbus 2021a. 

While this expression may look unwieldy, TDE discs have the fortunate property of being relatively cool, with their spectra peaking below the low band pass of X-ray telescopes, $kT  \ll 0.3$ keV. This means that X-ray observations of TDE  discs probe the quasi-Wien tail of the disc spectrum, a limit in which equation \ref{fullmodel} becomes analytically tractable. In Mummery \& Balbus (2021a) it was shown that equation \ref{fullmodel} can be integrated by performing a Laplace expansion about the hottest region in the disc, resulting in the following expression 
\begin{multline}\label{MB}
F_\nu(R_p, \T_p)  = \frac{4\pi \xi_1 h\nu^3}{c^2f_{\rm col}^4}\left( \frac{R_p}{D} \right)^2 \left(\frac{k \T_p}{h \nu} \right)^\gamma \exp\left(- \frac{h\nu}{k \T_p} \right) \\
\times \left[ 1 + \xi_2\left(\frac{k \T_p}{h \nu} \right) + \xi_3\left(\frac{k \T_p}{h \nu} \right)^{2}   +\, \dots \right], 
\end{multline} 
(see also Balbus 2014). Here we have defined $\T_p \equiv f_{\rm col} f_\gamma T_p$, where $T_p$ is the hottest temperature in the accretion disc. The radius $R_p$ corresponds to the image plane co-ordinate of this hottest region. The constant $\gamma$ depends on assumptions about both the inclination angle of the disc and the  disc's inner boundary condition, and is limited to the range $1/2 \leq \gamma \leq  3/2$. We note that $\gamma = 1/2$ for a vanishing ISCO stress disc observed precisely face on.  The positive constants $\xi_1, \xi_2$ and $\xi_3$ are all order unity, $\xi_1 \simeq 2.19, \xi_2 \simeq 3.50, \xi_3 \simeq 1.50$ (Mummery \& Balbus 2021a). Here `$\dots$' denotes higher order terms which scale like $\sim ({k \T_p}/{h \nu} )^{n }, n \geq 3$, which can be safely neglected. 

The flux from equation \ref{MB} depends only on the hottest temperature in the accretion disc, and not on contributions from other disc regions. This is an important result as TDE discs represent a class of accretion discs which are evolving, and are therefore not well described by steady state models. Standard disc models (e.g. the DISKBB model in XSPEC, Arnaud 1996) assume that the radial profile of the accretion disc is that of a steady state configuration. This enforces radial structure onto the solutions which may well not be present in reality, particularly at the earliest (and brightest) times. Equation (\ref{MB}) therefore represents a more agnostic approach to modelling TDE X-ray spectra, assuming only that the each disc radius emits like a colour-corrected blackbody, and that there exists some temperature maximum within the accretion disc. 

To compare this result to the single temperature blackbody function, consider its leading order $(k\T_p \ll h\nu)$ behaviour:
\beq\label{MB1}
F_\nu(R_p, \T_p)  \simeq \frac{4\pi \xi_1 h\nu^3}{c^2f_{\rm col}^4}\left( \frac{R_p}{D} \right)^2 \left(\frac{k \T_p}{h \nu} \right)^\gamma \exp\left(- \frac{h\nu}{k \T_p} \right)
\eeq
In this expression we observe key factors which differ from the single temperature blackbody model. A factor $2$ results from the disc, rather than spherical, geometry of the emission region. More importantly however, a factor $(1/f_{\rm col})^4$ results from disc {opacity} effects which, despite leaving the bolometric luminosity of the disc unchanged, strongly modify the overall scale of a discs Wien-tail flux. Finally, a factor $\xi_1({k \T_p}/{h \nu})^\gamma$ results from the fact that accretion discs have radial temperature profiles, as opposed to a single temperature. As $h\nu \gg k\T_p$, this factor further suppresses the overall scale of the disc emission. 

A single temperature blackbody is therefore missing two large factors which act to suppress the normalisation of the models flux. When fit to an observed X-ray spectrum, a single temperature blackbody will therefore correspondingly {\it underestimate} the radial size of the TDE emission region. Assuming that the exponential factors of the two models are comparable, and that they produce the same fluxes $(F_{\nu, \rm bb}/ F_{\nu, \rm disc} \sim 1)$, we have to leading order that
\beq
\left({R_{\rm bb} \over R_p }\right)^2 \sim 2\xi_1 \left({1 \over f_{\rm col}}\right)^4 \left(\frac{k \T_p}{h \nu} \right)^\gamma \sim 10^{-2}.
\eeq
Here we have assumed $k\T_p \sim 50$ eV, a typical value for TDE discs, and used the $f_{\rm col}(T)$ model of Done {\it et al}. (2012).

This simple result implies that approximating a disc spectrum with a single temperature blackbody will introduce an order of magnitude error into the inferred size of the disc. It is therefore of no surprise that the radial sizes inferred from TDE X-ray spectra are consistently found to be smaller than the event horizons of the black holes they occur around (e.g. Wevers {\it et al}. 2019, van Velzen {\it et al}. 2019, Stein {\it et al}. 2020, Cannizzaro {\it et al}. 2021) 

\section{Numerical demonstration}
In this section we verify two results, firstly, that the use of single temperature blackbody functions leads to order of magnitude under estimates of the physical sizes of TDE accretion discs. Secondly, that the disc fitting function (eq. \ref{MB}) accurately reproduces the sizes of accretion discs from their X-ray spectra.   To do this we generate simulated 0.3--10 keV X-ray spectra from fully relativistic numerical accretion disc models, and then fit these spectra with both a single temperature blackbody function and the disc fitting function.

\begin{figure}
 \centering
  \includegraphics[width=0.5\textwidth]{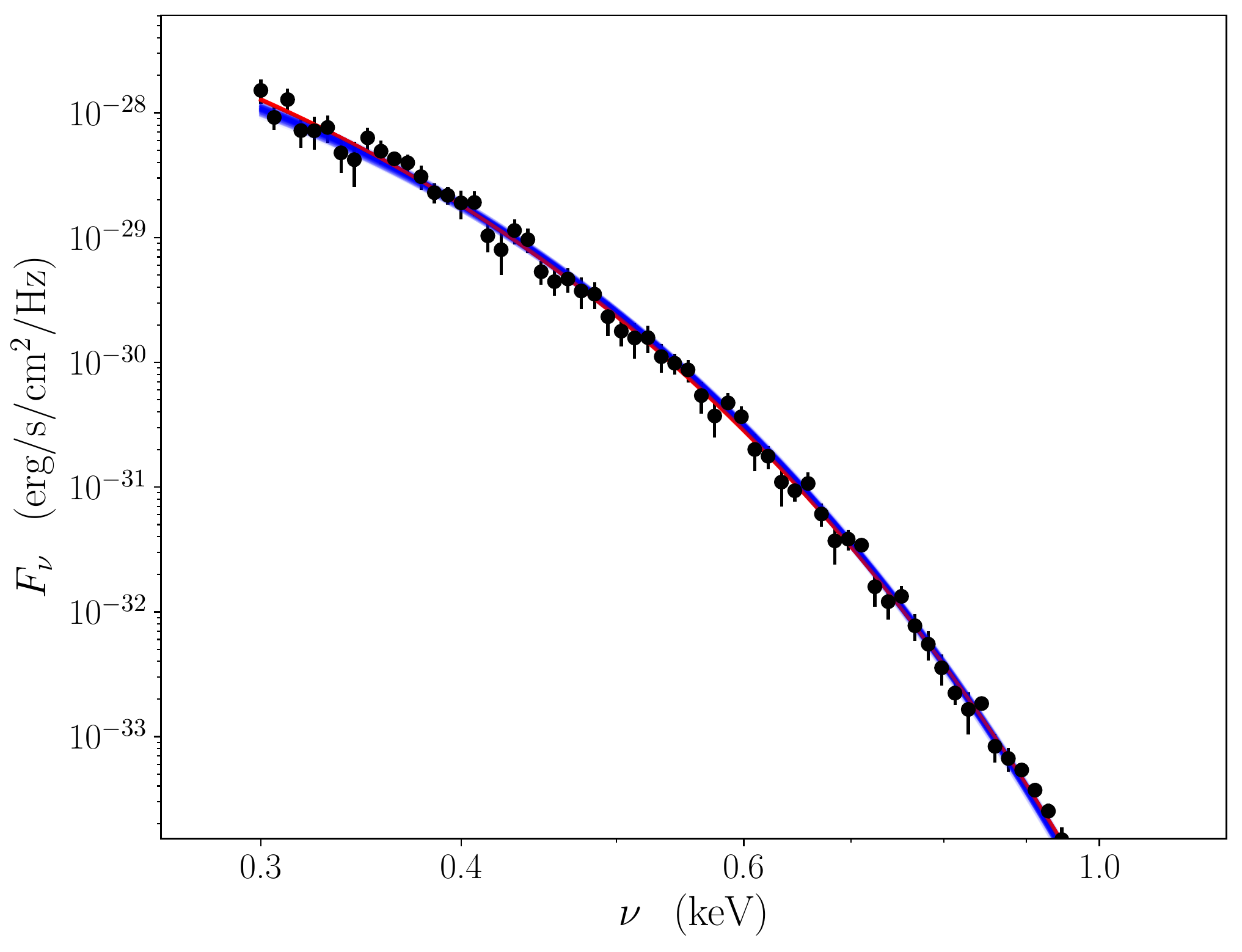} 
 \caption{Example samples drawn from the posterior distributions of the best fitting single temperature blackbody (blue curves), and best fitting disc model (red curves) fits to a simulated disc spectrum (black points).  Both functions produce perfectly adequate descriptions of the simulated data, but only the disc model provides useful information about the disc and black hole parameters of the system.   }
 \label{ExampleFits}
\end{figure}

`Fake' X-ray spectra are generated by solving the general-relativistic thin-disc evolution equation (Balbus \& Mummery 2018),  before ray-tracing the resulting disc emission profiles (black points, Figure \ref{ExampleFits}). We use the full temperature dependent colour-correction factor of Done {\it et al.} (2012). These models have, in the simplest modelling,  three free parameters: the black hole mass $M_{\rm BH}$, the disc mass $M_d$, and a disc $\alpha$ parameter which parameterises the disc turbulence. More complex models result from varying parameters like the black hole spin $a$, and the disc-observer inclination angle $\theta_{\rm obs}$. For a full description of the numerical techniques used to solve the relativistic disc equations and compute the discs resulting X-ray spectrum, see Mummery \& Balbus 2021a. 

\begin{figure}
 \centering
  \includegraphics[width=0.45\textwidth]{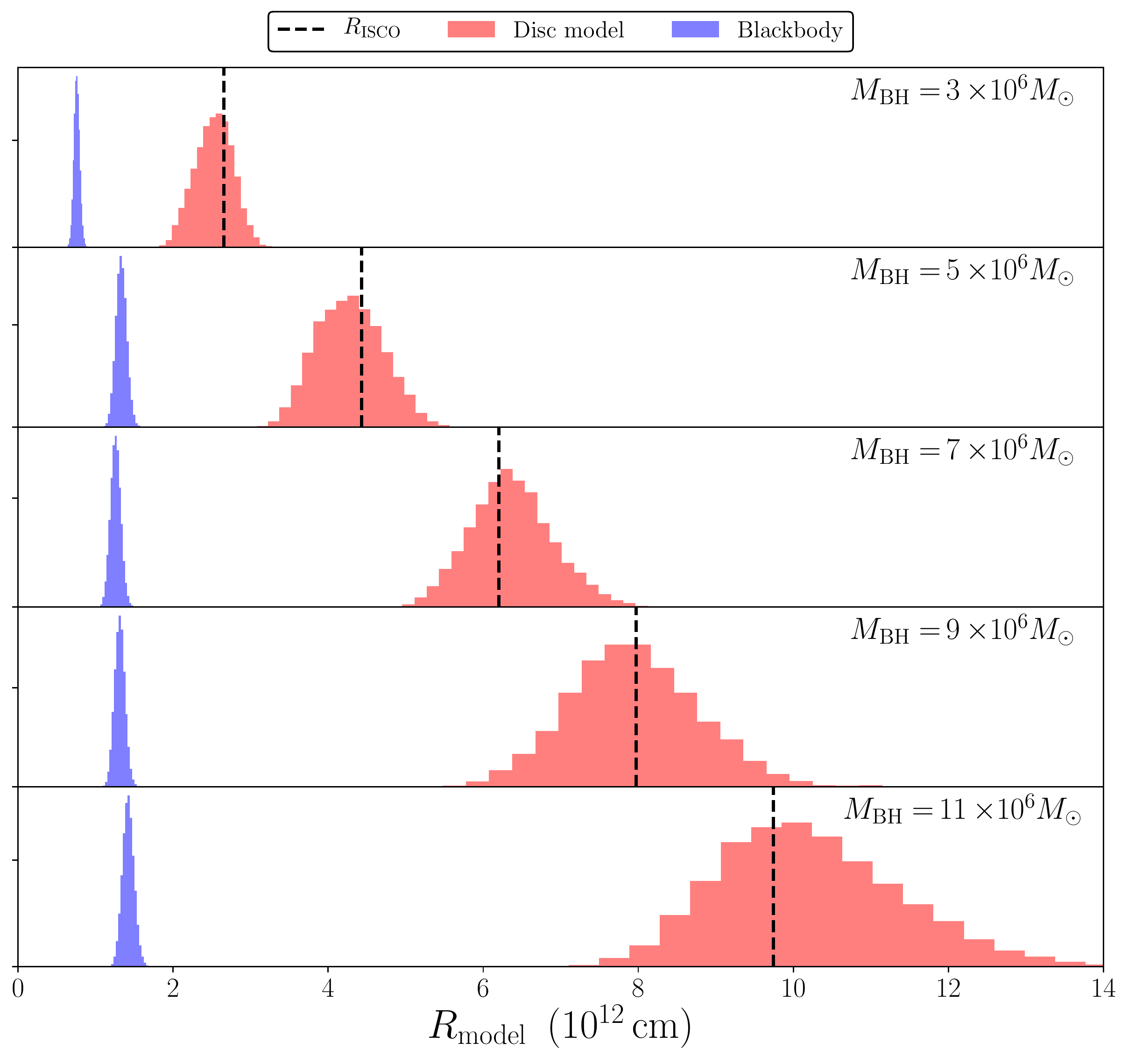} 
 \caption{The inferred posterior distributions of the two `emission size' fitting parameters, found using MCMC fits. The distribution of $R_{\rm bb}$ is plotted in blue, and $R_p$ in red.  The ISCO radius of each disc system is represented by a black dashed line. The black hole masses of each disc system are displayed in the upper right corner of each plot. The radial `size' inferred from fitting a single temperature blackbody to an observed disc spectrum bears no correlation with the actual size of that accretion disc. The radial sizes inferred from eq. \ref{MB} accurately reproduce the physical sizes of accretion discs. }
 \label{ModelComparison}
\end{figure}

A simple example of the deficiencies of the single temperature blackbody approach can be seen in Figure \ref{ModelComparison}.  Figure \ref{ModelComparison} shows the inferred posterior distributions of the radial size fitting parameters ($R_{\rm bb}$ and $R_p$) of the two spectral fitting models, found using MCMC fits (Foreman-Mackey et al. 2013). These models were fit to simulated X-ray spectra produced from five numerical disc models around black holes of different masses (denoted on plot). To simulate observation effects, random noise was added to each disc spectrum, at the level of $\sim 20\%$ of the flux in each frequency bin.   Each disc model was orientated face-on at an observer distance $D = 100$ Mpc, and the disc spectra were produced at the time when the bolometric luminosities of the evolving discs peaked. The discs around 5, 7 and 9 million solar mass black holes used $M_d = 0.5M_\odot$. To keep the disc luminosity sub-Eddington, the lowest black hole mass X-ray spectrum was produced with $M_d = 0.1 M_\odot$, while to produce observable levels of X-ray flux the largest black hole mass X-ray spectrum was produced with $M_d = 0.75 M_\odot$. All disc models used the canonical value $\alpha = 0.1$, and the black holes had spin parameters equal to zero. No prior constraints were placed on the fitting parameters, except $1/2 \leq \gamma \leq 3/2$. 

As can be clearly seen in Figure \ref{ModelComparison}, the radial `size' inferred from fitting a single temperature blackbody to an observed disc spectrum (blue histograms) bears no correlation with the actual size of the accretion disc which produces that spectrum. In fact, these best fitting blackbody radii do not even positively correlate with black hole mass, instead returning values $R_{\rm bb} \sim 10^{12}$ cm, independent of physical disc size. On the contrary, the radial sizes inferred from fitting eq. \ref{MB} to the simulated disc spectra (red histograms) accurately reproduce the physical sizes of the accretion discs (black dashed lines). 

It is important to stress that the point of this analysis is not that single temperature blackbodies produce particularly poor fits to the observed data, it is that the parameters inferred from such modelling do not correspond to any physical parameters of the system. This can be seen from Figure \ref{ExampleFits} which shows 150 samples drawn from the posterior distributions
of the best fitting single temperature blackbody (blue lines), and best fitting disc model (red lines). The simulated disc spectra are shown by black points, and correspond to the $M_{\rm BH} = 7\times10^6 M_\odot, M_d = 0.5M_\odot$ simulation. Both functions produce perfectly adequate descriptions of the simulated data, but only the disc model provides useful information about the disc and black hole parameters of the system. 

\section{Other sources of size underestimation } 
\subsection{Inclined disc systems}
\begin{figure}
 \centering
  \includegraphics[width=0.45\textwidth]{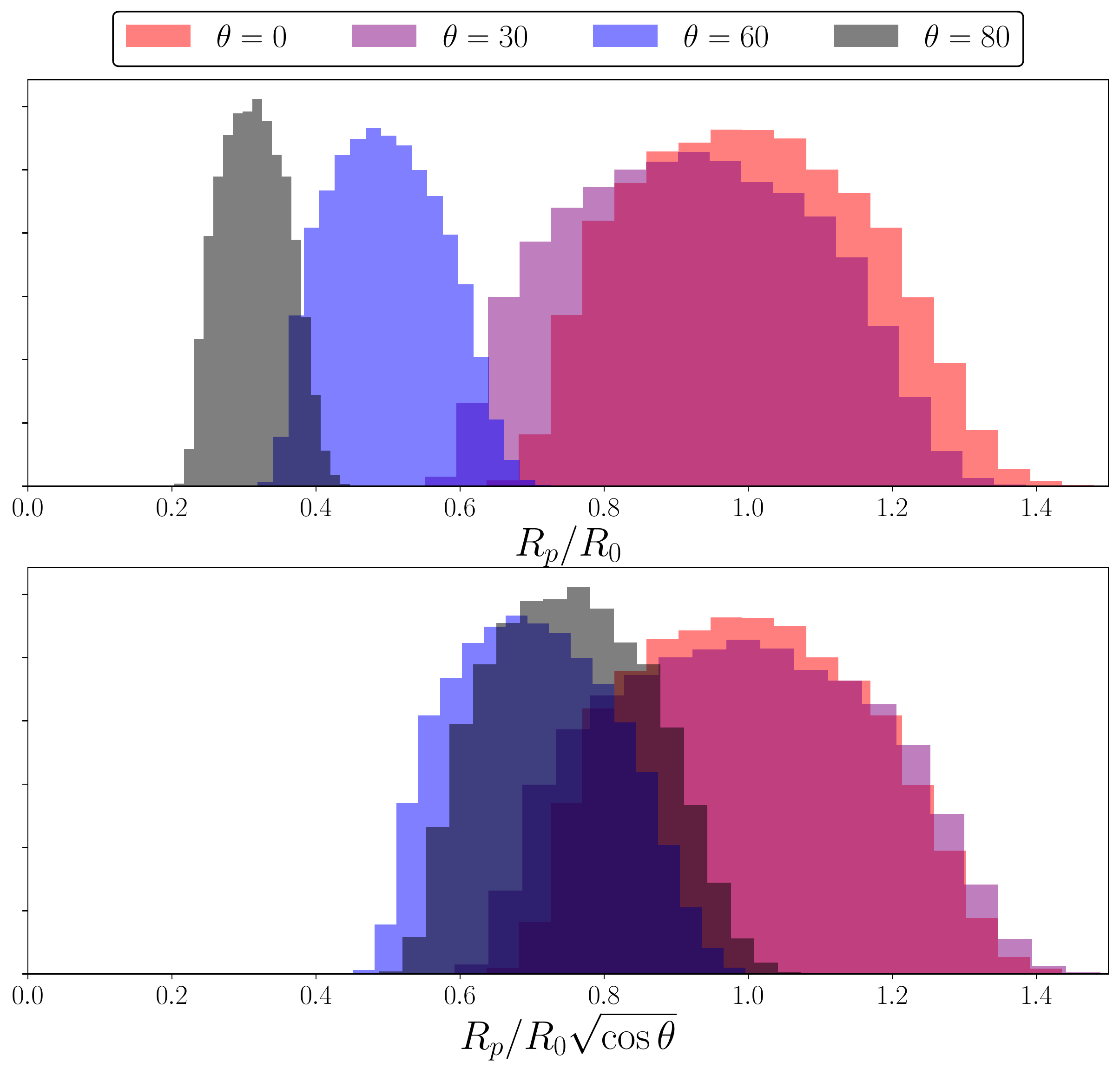} 
 \caption{The posterior distributions of the radial fitted parameter $R_p$ (equation \ref{MB}) fit to the observed 0.3-10 keV spectra of accretion discs inclined at an angle $\theta$ to the observer (inclination angle denoted in figure legend). The upper plot shows the distributions normalised by the best-fitting radius for a face on disc ($\theta = 0$), denoted $R_0$. The lower plot shows the distributions normalised by $R_0\sqrt{\cos\theta}$. Discs observed at an inclined angle will have inferred radii which are smaller than the physical radius of the disc. This correction factor can be as high as $\sim 3$ for large inclinations, and scales roughly as $\sqrt{\cos\theta}$. This analysis was performed for a black hole mass $M_{\rm BH} = 5 \times 10^6 M_\odot$, disc mass $M_d = 0.5M_\odot$ and $\alpha = 0.1$.     }
 \label{InclinedRadii}
\end{figure}

The accretion discs examined in the previous section were oriented face-on to the observer. Of course, in a real astrophysical system, TDE accretion discs will instead be oriented at some general angle $\theta$ from the observer. The best-fitting radial sizes inferred from an inclined disc system will be smaller than the physical disc radius (Figure \ref{InclinedRadii}). This is due to two principle effects. Firstly, simple Newtonian ray tracing introduces a factor $\cos(\theta)$ to the projected area of an accretion disc observed at an angle $\theta$. This then leads to an inferred radius $R_p = R_0 \sqrt{\cos\theta}$, where $R_0$ is the `true' disc radius. Secondly, inner-disc Doppler boosting of photons emitted from rapidly rotating disc material further decrease the inferred radii of inclined disc systems. Equation \ref{MB} demonstrates that the Wien-tail X-ray spectrum of an accretion disc is dominated by the region where the combination $\T = f_{\rm col} f_\gamma T$ is maximised. Both $f_{\rm col}$ and $T$ are axis-symmetric quantities, but the photon energy-shift factor $f_\gamma$ is, for large inclination angles, maximised in a small region where disc material is moving directly towards the observer. This rapidly rotating material has large Doppler blue-shifts, maximising $\T$, but correspond to only a small area of the accretion disc. 

As can be seen in the upper half of Figure \ref{InclinedRadii}, these two effects lead to a systematic underestimate of the radial size of an accretion disc when this disc is inclined to the observer. Fortunately however, this systematic effect can be approximately removed by identifying the fitted radius $R_p$ with $R_p = R_0 \sqrt{\cos\theta}$ (lower panel, Figure \ref{InclinedRadii}). It should be noted however, that due to Doppler blue-shifting this correction underestimates the required correction factor at large inclinations. Inclination angle effects cause at most a factor $\sim 3$ change in the inferred radii of a disc system.

\subsection{Local absorption}

\begin{figure}
 \centering
  \includegraphics[width=0.47\textwidth]{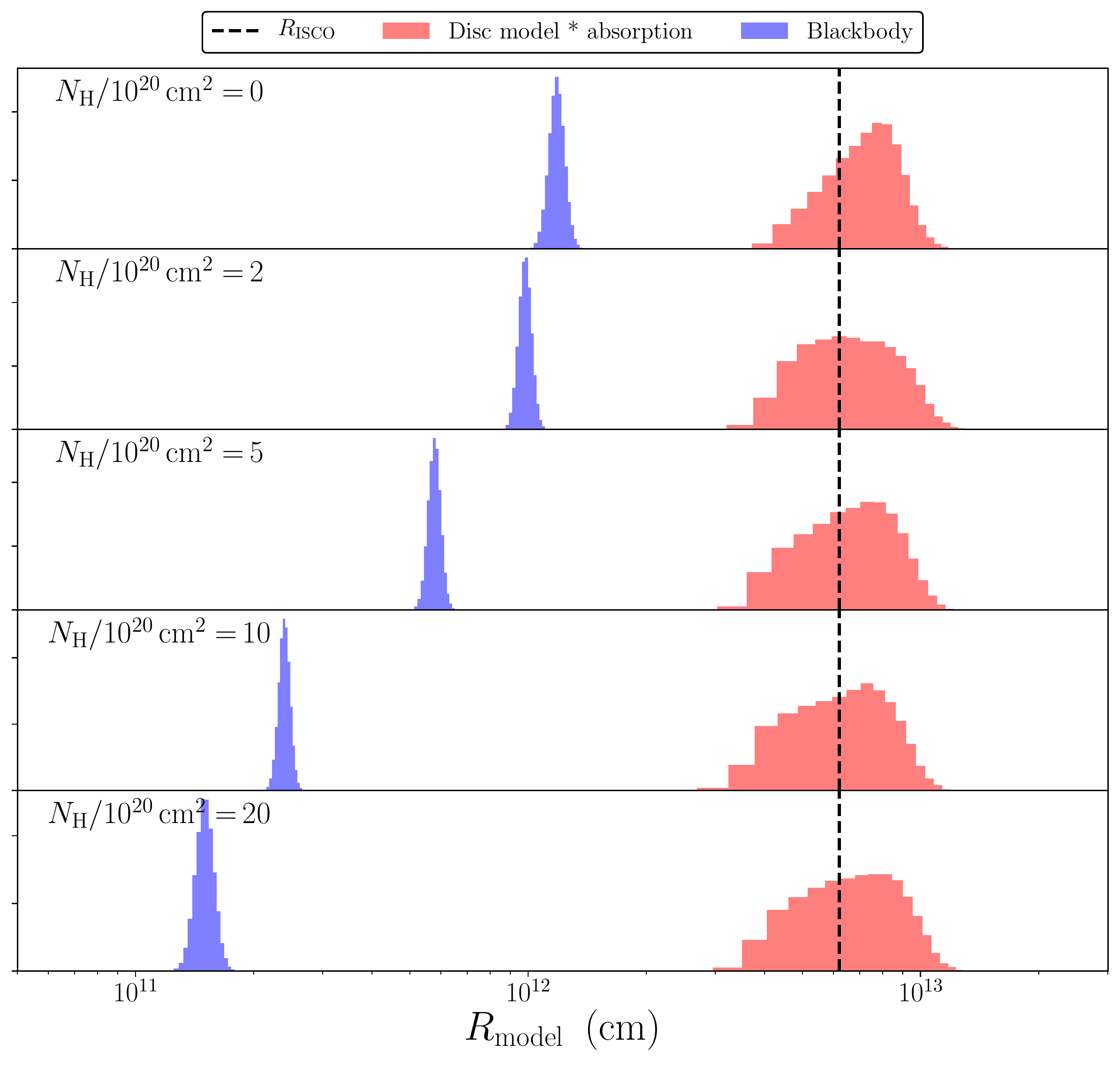} 
 \caption{The posterior distributions of the two `emission size' fitting parameters, found using MCMC fits to absorbed numerical disc spectra. The distribution of $R_{\rm bb}$ is plotted in blue, and the parameter $R_p$ of an absorbed disc model is shown in red.  The ISCO radius of the accretion disc  is represented by a black dashed line. The absorbing hydrogen column density of each disc spectrum is displayed in the upper left corner of each plot. The radial `size' inferred from fitting a single temperature blackbody to an absorbed disc spectrum differs hugely from the physical size of the disc. This analysis was performed for a black hole mass $M_{\rm BH} = 7 \times 10^6 M_\odot$, disc mass $M_d = 0.5M_\odot$ and $\alpha = 0.1$.    }
 \label{AbsModelComparison}
\end{figure}

A final factor which has been suggested as a potential cause of the deviations between the inferred and  physical radii of TDE systems is the presence of local absorption (e.g. Wevers {\it et al}. 2019). The idea here is that some stellar debris are expelled from the TDE disc system in the early stages of its evolution, which then obscure the accretion disc from the observer (Metzger \& Stone 2016). %, and X-ray photons emitted from the disc are reprocessed to Optical and UV frequencies (Metzger \& Stone 2016). 

As the ejected TDE material is likely to have been launched by radiatively driven winds produced around TDEs with particularly large bolometric luminosities, this effect will be particularly important for low black hole mass TDEs (Mummery 2021a).  The physical effect of absorption on an observed accretion disc spectrum is to introduce a photon-energy-dependent multiplicative factor which suppresses the spectrums amplitude. This takes the following form:
\beq\label{absorbed}
F_{\nu,  {\rm observed}} = \exp(-N_{\rm H}\, \sigma(E)) * F_{\nu, {\rm intrinsic}} ,
\eeq
where $N_{\rm H}$ is the Hydrogen column density, $\sigma(E)$ is the photon-energy-dependent scattering cross section per Hydrogen atom, and $F_{\nu, {\rm intrinsic}}$ is the un-absorbed flux, given by equation \ref{fullmodel}. The exact form of $\sigma(E)$ depends on the abundances of various elements in the absorbing medium, but can be well approximated analytically under certain reasonable assumptions (Morrison \& McCammon 1983). 

To examine the effects of local absorption on the inferred disc radii found from single temperature blackbody modelling we simulate  0.3-10 keV X-ray spectra from numerical disc models with varying absorbing column densities. The model for $\sigma(E)$ was taken from Morrison \& McCammon (1983). These absorbed disc spectra are then fit with both a single temperature blackbody, and an absorbed disc model (combining equations \ref{MB} and \ref{absorbed}, with $N_{\rm H}$ as a free fitting parameter). As can be seen in Figure \ref{AbsModelComparison},  fitting a single temperature blackbody model to a locally absorbed disc spectrum results in further large (up to a factor 10 for the largest column densities) deviations between inferred and physical disc radii. {This is in addition to the} intrinsic factor $\sim$ 10 underestimate inherent when using a single temperature blackbody function. 

\begin{figure}
 \centering
  \includegraphics[width=0.43\textwidth]{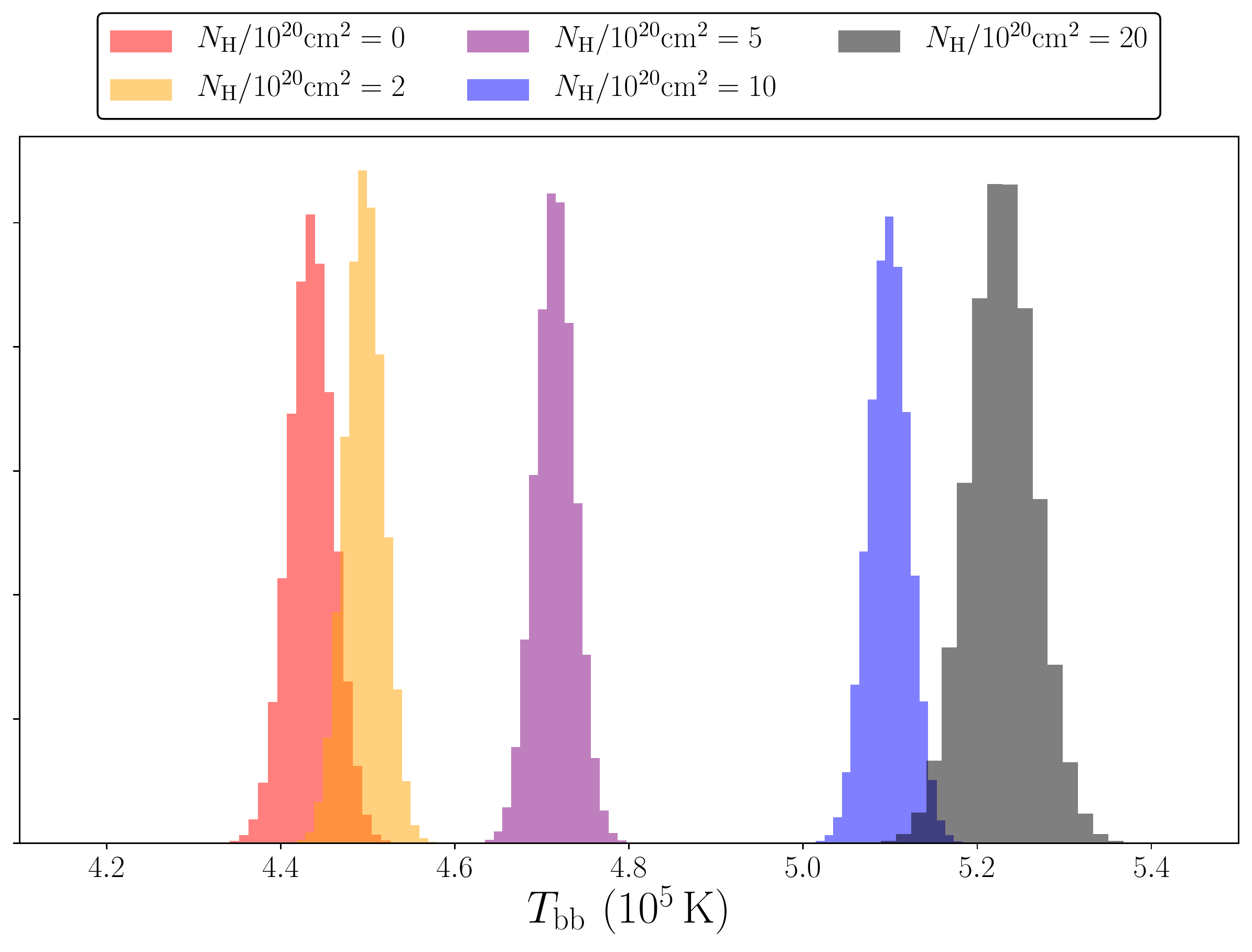} 
 \caption{The posterior distributions of the blackbody temperature, found using MCMC fits to absorbed numerical disc spectra for varying levels of absorbing Hydrogen column density. The absorbing hydrogen column density of each disc spectrum is displayed at the top of the plot. Disc spectra with higher levels of absorption result in higher inferred disc temperatures, due to additional curvature in the observed spectrum at low photon energies.    }
 \label{AbsTemp}
\end{figure}

 There are two reasons why higher absorption leads to further underestimation of disc emission radii.  Firstly, and most simply, absorption acts to suppress the amplitude of the observed X-ray emission (eq. \ref{absorbed}), which is directly translated to lower inferred radii. Secondly, the absorption cross section scales approximately as $\sigma \sim 1/E^3$ (but with various jumps due to atomic transitions, Morrison \& McCammon 1983), and thus introduce additional curvature in the low photon energy part of the spectrum. This additional curvature is then fit by a blackbody function with a higher temperature (Fig. \ref{AbsTemp}), which acts to further increase the amplitude of emission, thus requiring a further compensatory reduction in   blackbody radius. 

{However, i}t should be noted that fitting absorbed disc spectra using a single temperature blackbody function does not generally result in good fits to the data, particularly for large absorbing column densities $N_{\rm H} \gtrsim 10^{21} \, {\rm cm}^2$. Unabsorbed blackbody spectra tend to overestimate the flux at the lowest $E \sim 0.3$ keV photon energies (for the most absorbed disc spectra (lower panel, Figure \ref{AbsModelComparison}), the blackbody model was fit to the $0.4-10$ keV spectra, so that a good fit could be found).   {It is unlikely therefore that neglecting absorption has played a key role in the underestimation of TDE disc sizes in the literature, because good fits are generally found with blackbody modelling. In fact, a signature of significant intrinsic neutral absorption is  rarely detected in high-quality TDE spectra (Saxton {\it et al}. 2021). Undetected neutral absorption may however cause further underestimation of blackbody radii in the lowest quality TDE X-ray spectra, where the signal-to-noise is insufficient to disentangle the absorption and emission components.}

As can be seen in Figure \ref{AbsModelComparison},  the correct disc radius can be recovered from an absorbed X-ray spectrum by treating the absorbing column density $N_{\rm H}$ as a free parameter in a disc model fit  (red histograms).  {Modelling of TDE X-ray spectra should contain an absorption component, in addition to a disc model, to model any intrinsic neutral absorption which may exist in the system (e.g. the ZTBABS model (Wilms et al. 2000) in XSPEC).}

\section{Conclusions}
In this paper we have demonstrated that modelling TDE X-ray spectra with a single temperature blackbody function leads to the systematic underestimation of the sizes of their accretion discs by as much as an order of magnitude. In fact, the `blackbody radii'  found with this method do not even positively correlate with the physical size of an accretion disc (Figure \ref{ModelComparison}). Coupled with {possible (generally smaller)} corrections factors from inclination angle and local absorption effects, this work explains why various authors infer X-ray emission radii which are smaller than the event horizons of TDE black holes. 

The size and temperature of TDE accretion discs are important observational parameters as they encode information about both the black hole and disc at the heart of a TDE. By using the disc model presented in this paper (eq. \ref{MB}), these two parameters can now be accurately inferred from future X-ray observations of TDEs.  

\section*{Acknowledgments} 
I am grateful for conversations with Adam Ingram \& Thomas Wevers, and to the reviewer, Richard Saxton, for helpful comments.
%\section*{Data accessibility statement}
%An XSPEC implementation of the disc fitting function is available at github.com/andymummeryastro/TDEdiscXraySpectrum.
\label{lastpage}

\end{document}